\documentstyle[12pt,epsf]{article}
\begin{document}
\begin{flushright}
TAUP-2347-96\\
hep-th/9607028
\end{flushright}
\baselineskip 22pt
\newcommand{\be}{\begin{equation}}
\newcommand{\ee}{\end{equation}}
\newcommand{\leqx}{\,\raisebox{-1.0ex}{$\stackrel{\textstyle <}
{\sim}$}\,}
\newcommand{\w}{Schwarzschild $\:$}
\newcommand{\geqx}{\,\raisebox{-1.0ex}{$\stackrel{\textstyle >}
{\sim}$}\,}
\newcommand{\th}{\theta}
\newcommand{\va}{\varphi}
\newcommand{\de}{\Delta}
\newcommand{\x}{\tilde{x}}
\newcommand{\bl}{\hspace{-.65cm}}
\newcommand {\s}
[1] {\mid \! \! {#1} \rangle}
\begin{center}
{\bf Is the   Black Hole Complementarity principle really
 necessary?}
\\N.Itzhaki \footnote{Email Address:sanny@post.tau.ac.il}
\\Raymond and Beverly Sackler Faculty of  Exact Sciences
\\School
 of Physics and Astronomy
\\Tel Aviv University, Ramat Aviv, 69978, Israel
\end{center}
\begin{abstract}
We show that the S-matrix ansatz implies a semi-classical
 metric such that a freely falling test particle will  not
 cross the horizon in its proper time.
 Instead of reaching the singularity it will reach ${\cal I^{+}}$.
\end{abstract}
%\newpage

One of the most interesting questions that one has to confront
if one adopts 't Hooft S-matrix ansatz \cite{th90} is whether 
or not a
freely falling observer can cross the horizon in   
  his way to the singularity.
On the one hand, the curvature is small at the horizon, of the
order of $\frac{1}{M^2}$ \footnote{In units where $G=\hbar =c=1$.},
 and at least classically there is
nothing special at the horizon for a freely falling observer.
Furthermore, a freely falling observer cannot detect the Hawking 
temperature since for such an observer
the  Hawking radiation is   part of the vacuum fluctuations.
On the other hand, according to the S-matrix ansatz the 
information is encoded in the Hawking radiation due to strong 
gravitational interactions that take place just outside
 the horizon.

This conflict lead  Susskind, Thorlacius and Uglum to suggest 
 the black hole complementarity principle \cite{sus}
 which can be formulated as follows:

\bl $\bullet$ From the point of view of an external observer the region
just outside the horizon (stretched horizon) acts like a very
 hot membrane which absorbs thermalizes and emits any 
information that falls to the black hole.

\bl $\bullet$ From the point of view of a freely falling observer
there is nothing special at the horizon so a freely falling observer
  can  cross the horizon in his way to the singularity.  

At first sight, this principle seems to be
inconsistent. But a more careful analysis of   some  
 Gedanken experiments shows that  
 the black hole complementarity principle may be  consistent 
\cite{sus}.
  
In this letter we do not consider the consistency of the black hole
complementarity principle but we show that it is not necessary,
 namely that the S-matrix ansatz is complete by itself.
In particular, we show  that the S-matrix ansatz implies 
such a  gravitational back-reaction that a freely falling
 observer will not cross the horizon even in his  proper time.
This implies that  the 
 Penrose diagram describing gravitational collapse  
 has the same topology as Minkowski space.  
Such a Penrose diagram was already suggested  in
\cite{th93,eng} using the weak value \cite{aha} of the metric as
defined by an external observer.
Our point is that we  do not consider the weak value but the mean
value as implied by the S-matrix ansatz.
Therefore our results hold also for an infalling observer.

Classically, a freely falling observer can cross the horizon
in a finite amount of proper time.
Quantum mechanically, the black hole radiates so  a freely falling observer 
is moving along a geodesic in the metric induced by the black hole and
the Hawking radiation.
In order to find the geodesics one should be able to calculate the 
mean energy-momentum tensor of the radiation, $T_{\mu\nu}^{rad}$, and to
take it  as a source to the Einstein's equations in a \w background.
However,  in field theory 
$T_{\mu\nu}^{rad}$ diverges and should be  renormalized.
Unfortunately, the Einstein-Hilbert Lagrangian is non-renormalizable.
Nevertheless, under some ``reasonable'' assumptions one can regularize and
renormalize  $T_{\mu\nu}^{rad}$ where for
 simplicity the star in 
is taken to be a thin spherically symmetric shell and the centrifugal
barrier is neglected (only the s-wave sector is taken into account) \cite{un}.
The ``reasonable'' assumption   is that the metric and 
$T_{\mu\nu}^{rad}$ are regular for a freely falling observer.
Naturally, this assumption leads to  a metric  which allows a freely falling
observer to cross the horizon.
The mean energy-momentum tensor of a spin-less massless field that
 was found in \cite{un} is
\begin{eqnarray}
  \label{un}
 4\pi r^2 T_{uu}^{rad}&=&\frac{1}{12\pi}\left[-\frac{M}{2r^3}
(1-\frac{2M}{r})-\frac{M^2}{4r^4}\right]+\frac{\pi }
{12 (8\pi M)^2}\\ \nonumber
4\pi r^2 T_{vv}^{rad}&=&\frac{1}{12\pi}\left[ -\frac{M}{2r^3}
(1-\frac{2M}{r})-\frac{M^2}{4r^4}\right]\\ \nonumber
4\pi r^2 T_{uv}^{rad}&=&-\frac{M}{12\pi 2r^3}(1-\frac{2M}{r}),
\end{eqnarray}

The physical origin of this energy-momentum tensor is that
pairs of ``particles''  are created in the region above the horizon one
with negative energy which falls into the black hole, and one with positive
energy which reaches ${\cal I^+}$.
Note that for any $r$, 
$4\pi r^2 (T_{uu}-T_{vv})=\frac{\pi}
{12 (8\pi M)^2}$
 which is  the thermal flux of the Hawking radiation,
up to a numerical constant due to the fact that the centrifugal
barrier was dropped.

Notice further that near the horizon the mass evaporation  is due to a 
  negative
$T_{vv}^{rad}$ and not to a positive $T_{uu}^{rad}$ (at $r=2M $ one gets
 $T_{uu}^{rad}=0$) .
Thus in this scenario 
at least  semi-classically there are no strong gravitational
interactions near the horizon supporting  Hawking's picture \cite{haw} .
The reason is that in this scenario near the horizon all the energy
momentum flux is
in the $v$ direction .
Namely, $T_{vv}$ is positive for ingoing light particles
and negative for Hawking radiation and all other components of
$T_{\mu\nu}$ vanish,
so $T_{\mu\nu}^{rad}$ and  $T_{\mu\nu}^{in}$ are parallel 
hence they do not interact.
As a result there is an {\em exact} solution to Einstein's equation near the
horizon of a black hole which is constructed out of ingoing light like
flux including the back reaction of Hawking particles.
 The exact solution can be expressed by the ingoing
Eddington-Finkelstein coordinate (the Vaidya solution):
\be ds^2=-(1-\frac{2M(v)}{r})dv^2+2dvdr+r^2 d\Omega^2\ee
 where 
\be M(v)=\int_{}^{v}dv^{'}(T_{v^{'}v^{'}}^{rad}+T_{v^{'}v^{'}}^{in}).\ee
The ingoing null geodesics in this metric are $v=const$, so it is clear
that in this scenario a freely falling observer
 can cross the horizon (for a recent review see \cite{ser}).
But it is important to emphasis that in fact that result
 was assumed by the regularity condition at the horizon.

%{\bf S-matrix ansatz and backreaction}
This scenario cannot coexist 
with   the S-matrix ansatz.
The reason is that if the information is encoded in Hawking radiation 
 then there should be strong interactions
between  ingoing and outgoing particles.
Causality implies that interactions
taking place  behind the horizon will not affect the final
 state of the radiation and interactions which occur at large distances from
the horizon are between regular particles (no Planckian
 energies) so their effect is   too weak.
 Therefore the S-matrix ansatz implies that strong interactions
 should take place just outside the horizon unlike Hawking's scenario.   

In \cite{th90} an approximation to the S-matrix was suggested.
The approximation is based on the classical 
gravitational interaction between two light particles (the 
gravitational shock wave) and the WKB approximation.
The gravitational field of a massless particle in Minkowski space is
described by  the line element
\be ds^2 =-dU(dV+4p\ln (\frac{\x^2}{M^2})\delta (U-U_0)dU)+dx^2+dy^2\ee
where $\x ^2=x^2+y^2$.
The massless particle moves in the $V$ direction with constant
 $U=U_0,\; \x=0$ and momentum $p$ \cite{sex}.
The  effect of this metric on null geodesics is
a discontinuity $\delta V$ at $U=U_0$ \cite{th84}
\be \delta V(\x )=-4p\ln (\frac{\x ^2}{M^2} )\ee
Using the WKB approximation one can find \cite{th90} that 
up to an overall phase only one S-matrix agrees with Eq.(4)
\be  \langle p_{out}(\x^{'} )\mid p_{in}
(\x )\rangle=
N\exp \left(4i \int d^2\x d^2\x ^{'}p_{out}
(\x ) f(x^{'}, x) p_{in}(\x^{'})\right),\ee
where $p_{out}(\x^{'})$ and $p_{in}
(\x )$ are the momentum distributions of the out-state and in-state and
   $f(x^{'}, x)$ is the Green function on the horizon 
$\ln (\frac{(\x-\x^{'})^2}{M^2} )$.
Notice that Eq.(6) is symmetric under time reversal.
In fact the basic idea of quantum mechanics and hence of 
the S-matrix ansatz is to  treat the
out-state and in-state on an equal footing .\footnote{Treating the in-state and
  outstate symmetrically is also the key in the weak value approach
 \cite{th93,eng}.}
Since for ingoing light flux  $T_{uu}^{in}=T_{uv}^{in}=0$, time
reversal implies that for outgoing  radiation $T_{vv}^{rad}=T_{uv}^{rad}=0$.
Therefore, the S-matrix ansatz implies that

\begin{eqnarray}
\label{sm}
4\pi r^2 T_{uu}^{rad} &=& \frac{\alpha }{M^2}\\ \nonumber
4\pi r^2 T_{vv}^{rad} &=& 0 \\ \nonumber
4\pi r^2 T_{uv}^{rad} &=& 0
\end{eqnarray}
where $\alpha$ depends on the number of radiated fields and their spin
\cite{page}. 
We are now in a position to find the geodesics according to
 the S-matrix ansatz.
We shall obtain the same result using two alternatives derivations.
The first approach is to use the Rindler approximation to the \w metric
near the horizon,
\be ds^2=dUdV+dX_i dX_i \ee
Rindler and \w coordinates are related by
\begin{eqnarray}
\rho ^2=UV \\ \nonumber
t=2M\ln (U/V)
\end{eqnarray}
where $\rho $ is the invariant distance from the horizon in a fixed
\w time
\be \rho=\int_{2M}^{r}dr \sqrt{g_{rr}}\approx \sqrt{8M(r-2M)}\ee 
The effect  of the gravitational shock wave
 of one outgoing Hawking particle at $V=V_0$ on an ingoing
 test particle is a discontinuity 
$\delta U\approx 1/V_0$.
 As a result $\rho $ increases (see Figure~1).
If there were only one Hawking particle then the test particle would
have still cross the horizon (at a later time than it would have in the
absence of the outgoing particle).
But, before the test particle crosses the horizon it 
crosses the shock waves of all  the Hawking particles.
By the time it has crossed all   the shock waves 
 the black hole has already evaporated completely 
so there is no mass left to form a black hole and a horizon.
 To obtain this conclusion in a rigorous way
 we can use Eqs.(4,7) and $u=4M\ln (U/4M)$ 
to find  that the back-reaction of Hawking flux on
the metric near the horizon is 
\be ds^2=dU dV+\frac{a}{U^2}dU^2+dX_i dX_i,\ee
where $a=16\alpha $.
The null geodesics lines can  be found for radial trajectories 
\begin{eqnarray}
  \label{geo}
 U=U_0\\ \nonumber
V=\frac{a}{U}+V_0
\end{eqnarray}
where $V_0$ and $U_0$ are positive constants.
Therefore, the outgoing trajectories are the same 
as in Rindler space but  the  ingoing
trajectories are such that   
\be \rho ^2=UV =a+U V_0\geq a\ee 
So an ingoing particle will  eventually float at distance $\sqrt{a}$ above the
horizon instead of crossing it.

\begin{figure}
\begin{picture}(300,300)(0,0)
\mbox{\epsfxsize=100mm \epsfbox{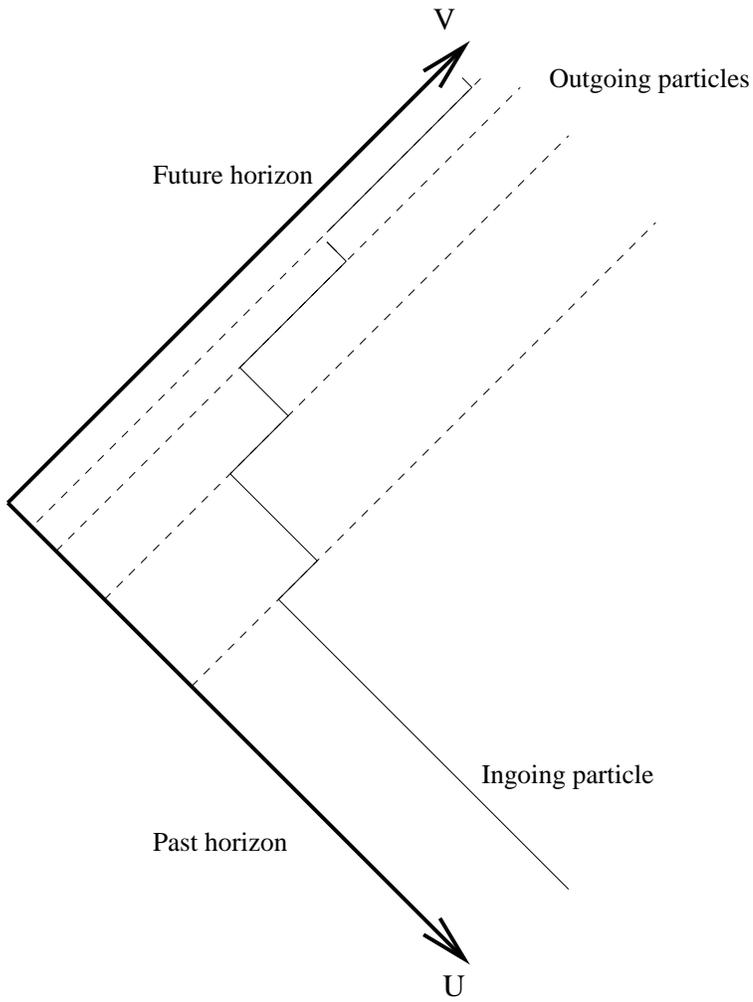}}
\end{picture}
\caption{The effect of the shock wave of the Hawking radiation
   on an in going particle.}
\end{figure}

The inertial coordinate system of the  test particle can be found 
by making a coordinate change 
\begin{eqnarray}
 V^{'}= V-\frac{a}{U}\\ \nonumber
U^{'}=U
\end{eqnarray}
to get
\be ds^2=dU^{'}dV^{'}+dX_i dX_i \ee
The horizon in this metric  is at $\rho^{'^{2}}=V^{'}U^{'}=-a$ 
 but this is a fictitious horizon since   when the 
ingoing particle  reaches $U^{'}=0$ all Hawking particles were
already emitted
so there is no longer a black hole.  
We see therefore that
 the test particle which started at a flat space ${\cal I^{-}}$
will end up in  flat space and not at the singularity
 (Figure 2a) though along the trajectory there is a region with
 large curvature.

\begin{figure}
\begin{picture}(300,300)(0,0)
%\vspace{5mm}
\hspace{18mm}
\mbox{\epsfxsize=120mm \epsfbox{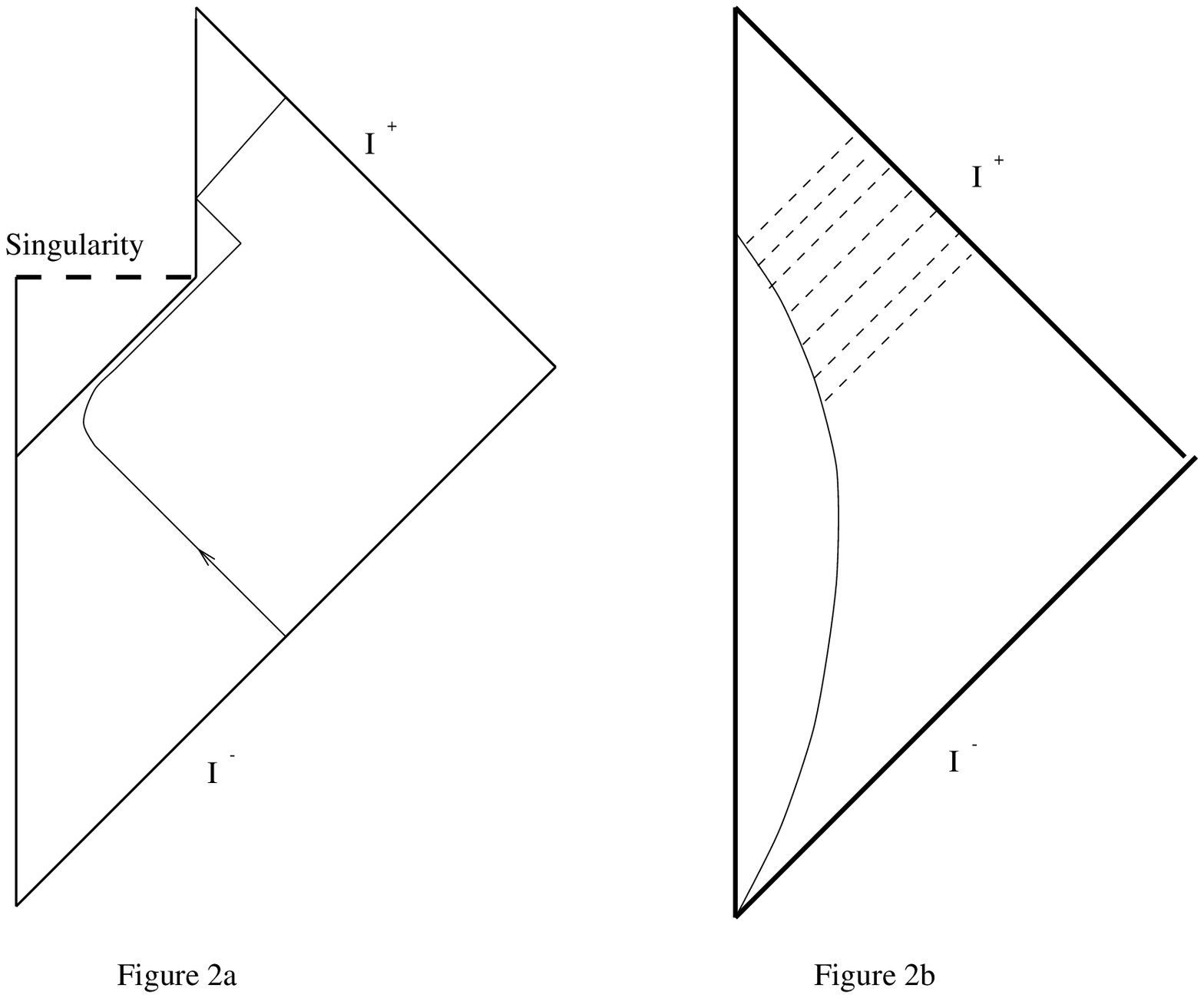}}
\end{picture}
\caption{The trajectory of an ingoing light particle is such that it
 will not cross the horizon but  will reach ${\cal I^{+}}$.
Therefore, the Penrose diagram describing a collapsing star has a
trivial topology.}
\end{figure}
An alternative way to reach the same result is to use the outgoing 
 Eddington-Finkelstein coordinates.
In the region where  there is no infalling matter but
only  Hawking radiation Vaidya metric is  an exact solution to
 Einstein's equation including
Hawking radiation where $T_{\mu\nu}^{rad}$ is given according to the
S-matrix ansatz (Eq.(7)). 
\be ds^2=-(1-\frac{2M(u)}{r})du^2-dudr+r^2d\Omega^2,\ee 
where 
\be M(u)=M-\int_{u_0}^{u}du T_{uu}\ee 
 $u_0$ is the time at which  Hawking radiation started.
The ingoing radial null geodesic equation is 
\be \frac{dr}{du}=-(1-\frac{2M(u)}{r})\ee
Eventually the test particle will be near the horizon and it is
helpful to define 
\be \delta(u)=r(u)-2M(u)\ee
to get
\be \frac{d\delta}{du}\approx \frac{\delta}{M}+2\frac{dM}{du}=
\frac{\delta}{M}-\frac{2\alpha}{M^2}\ee
So asymptotically $\delta(u)= \frac{2\alpha }{M(u)}$.
This means that the invariant distance from the horizon is $\rho
=\sqrt{a}$ which is the same result as Eq.(14).
Note that  the exact form of $M(u)$ is not important to reach the
conclusion that the test particle will not cross the horizon.
 The exact form of $M(u)$ is only important to find the invariant
 distance from the horizon at which the null geodesic will float. 
Therefore, fluctuations of $T_{uu}$ cannot change the result that
test particle cannot cross the horizon. 
Clearly, both  Eq.(11) and Eq.(17) are not a good approximation to 
the metric in the region where there are ingoing particles. 
Nevertheless, the above discussion is a strong indication that the S-matrix
 ansatz implies such a back reaction that the  
 Penrose diagram of a star which collapses to form a black
hole has the same topology as Minkowski space (see Figure 2b).

\vspace{1.5cm}

I am grateful to S.Massar for his suggestion to use Vaidya solution
and for helpfull discussions. 
I would also like to thank  Y. Aharonov, A. Casher, F. Englert and R.
Parentani for useful discussions.

\end{document}